\documentclass[a4paper.12pt]{article}
\title{Computing Biomolecular System Steady-states} 

\usepackage[noblocks]{authblk}
\author{Peter J. Gawthrop\footnote{Corresponding author. \textbf{peter.gawthrop@unimelb.edu.au}}}
\affil{
  Systems Biology Laboratory,
  Department of Biomedical Engineering,
  Melbourne School of Engineering,
  University of Melbourne,
  Victoria 3010, Australia.
   }

\usepackage{a4wide}

\usepackage[numbers]{natbib}
\usepackage{amsmath,amssymb,amscd,amstext,mathtools,extarrows,centernot}
\newcommand{\asinh}{\text{ asinh }}
\usepackage{chemformula}

\newcommand{\NN}{N}
\newcommand{\Nf}{{\NN^f}}
\newcommand{\Nr}{{\NN^r}}

\newcommand{\Ncd}{\NN^{cd}}
\newcommand{\VV}{V}

\newcommand{\vv}{v_0}
\newcommand{\svv}{\vv^\star}

\newcommand{\svveq}{\vv^{eq}}
\newcommand{\vp}{\vv^+}
\newcommand{\vm}{\vv^-}

\newcommand{\VN}{\phi_N}
\newcommand{\sPhi}{\Phi^\star}
\newcommand{\skappa}{\kappa^\star}

\newcommand{\skappaeq}{\kappa^{eq}}
\newcommand{\kappabar}{\bar{\kappa}}

\usepackage{mhchem}

\usepackage{amsmath,amssymb,amscd,amstext,mathtools,extarrows}

\newcommand{\hp}{{\cdot}}         
\newcommand{\Std}{\ominus}
\newcommand{\std}{\oslash}

\newcommand{\dX}{\dot{X}}
\newcommand{\Gcm}{G_{cm}}
\newcommand{\phicm}{\phi_{cm}}
\newcommand{\lb}{\left (}
\newcommand{\rb}{\right )}
\newcommand{\reacul}[2]{
  {\; \xrightleftharpoons[#2]{#1} \;}
}

\newcommand{\reacu}[1]{
  \reacul{#1}{}
}

\usepackage{siunitx}
\newcommand{\BG}[1]{\text{\sffamily\textbf{#1}}}

\newcommand{\C}{\BG{C }}

\newcommand{\one}{\BG{1 }}
\newcommand{\zero}{\BG{0 }}

\renewcommand{\Re}{\BG{Re }}

\newcommand{\BGL}[2]{$\BG{#1}$:$\mathbf{#2}$} 

\newcommand{\BC}[1]{\BGL{C}{#1}}

\newcommand{\BRe}[1]{\BGL{Re}{#1}}

\usepackage{listings}

\usepackage{graphicx,subfigure,fancybox,color}
\newcommand{\FigWidth}{0.75}
\newcommand{\Fig}[2]{
  \includegraphics[width=#2\linewidth]{#1.pdf}
   \label{subfig:#1}
}

\newcommand{\SubFig}[3]{
  \subfigure[#2]{
    \includegraphics[width=#3\linewidth]{#1.pdf}
    \label{subfig:#1}
  }
}

\usepackage{hyperref}
\hypersetup{colorlinks=true,citecolor=blue,linkcolor=blue,urlcolor=green}
\usepackage{url}
\usepackage{doi}

\begin{document}
\maketitle
\begin{abstract}
  A new approach to computing the equilibria and steady-states of
  biomolecular systems modelled by bond graphs is presented. The
  approach is illustrated using a  model of a biomolecular cycle
  representing a membrane transporter and a model of the mitochondrial
  electron transport chain.
\end{abstract}


\section{Introduction}
\label{sec:introduction}
There are many ways of deriving the dynamical equations describing the
time course of flows and concentrations in biomolecular systems
\citep{Bea11,Bea12}.  As physical systems, biomolecular systems are
subject to the laws of thermodynamics \citep{BeaQia10,AtkPau11} and
number of such \emph{energy-based} approaches have been developed
including the bond graph of \citet{OstPerKat71,OstPerKat73} further
developed by \citet{GreCel12} and
\citet{GawCra14,GawCra16,GawCra17}. The bond graph approach is
particularly appropriate for systems which combine different physical
domains and thus are particularly appropriate for electrogenic
biomolecular systems \citep{GawSieKam17,Gaw17a}. In common with other
approaches, the bond graph approach can be used to derive ordinary
differential equations (ode) describing the time evolution of the
chemical species, electrical charges and chemical and electrical
flows.

The basic ideas of bond graphs can be found in textbooks
\citep{Bor10,KarMarRos12} and a tutorial for engineers \citep{GawBev07}.
Appendices A \& B of a recent paper \citep{GawCra17} provide a
tutorial introduction to bond graphs in the context of biomolecular
systems; in particular, the stoichiometric matrix describing the
biomolecular system can be automatically derived from the bond graph. In
this paper, both the bond graph and relevant stoichiometric matrices
are stated but the derivation is omitted to save space and to focus
attention on the novel ideas of the paper.

In some cases, it is of interest to find values corresponding to a
situation where the amount of each chemical species and electrical
charge is constant; this is termed a \emph{steady-state} of the
system. System steady-states correspond to constant chemical and
electrical flows; in the special case the such flows are zero,
the system is said to be in \emph{equilibrium}.
For example, study of the energetic behaviour of metabolic
biomolecular systems at constant flow leads to conclusions about
efficient enzyme utilisation \citep{ParRubXu16}.

A number of approaches to finding the steady-state of biomolecular
networks have been suggested including simulation, flux balance
analysis~\citep{OrtThiPal10} and the diagram methods of
\citet{KinAlt56} and \citet{Hil89}.
 
The simulation approach is exemplified by \citet{BazBeaVin16} who
state that ``For the steady state analysis, the model was simulated
until the absolute values of the state variable derivatives were less
than $10^{-10}$ ...''.
A similar approach is to simulate starting from equilibrium with
slowly-varying chemostats followed by an algebraic solver for each
simulation time point starting with the simulated state
\citet{GawCra17}. Such approaches can be numerically challenging and time
consuming.
%
Flux balance analysis (FBA) \citep{OrtThiPal10} is based on the system
stoichiometric matrix but  ignores reaction kinetics.
Essentially, the system
stoichiometric matrix constrains all steady-state flows
$\VV$ to lie within the right null subspace of the chemodynamic
stoichiometric matrix $\Ncd$. 
These constraints are not enough to deduce the steady-state flows
$\VV$. Specific values for $\VV$ are found by adding extra constraints
together with a quadratic (in the flows) cost function and applying
linear programming.
Energy considerations are not directly addressed, but can be added on
\citep{HenBroHat07}.
The diagram method of \citet{KinAlt56} and \citet{Hil89} can be used
to generate explicit rate equations for reaction networks using, for
example, the software described by \citet{QiDasHan09}.

A general bond graph approach to steady state analysis using causality
concepts was introduced by \citet{Bre84b} and extended by
\citet{BidMarSca07} and \citet{GonGal11}. In contrast, the approach of
this paper is specifically tailored to the special properties of
biomolecular system bond graphs: no inertial components, conserved
moieties leading to initial condition dependent steady-states and
particular non-linearities involving logarithms and exponentials.
The special bond graph junction structure of biochemical systems is
noted elsewhere \citep{Gaw17}. In particular, in a similar fashion to
\citep{Bre84b}, the application of derivative causality reveals key
stoichiometric properties such as conserved moieties and flux
pathways. 

The approach to steady-state analysis proposed here is based on two
ideas: using the \emph{Faraday equivalent potential} \citep{Gaw17a} to
unify chemical and electrical potential and using the positive-pathway
approach \citep{GawCra17} to represent the steady-state flow pathways.

The approach suggested here is like the FBA approach insofar as it
uses pathway analysis of steady-state flows, and like the simulation
approach in that it uses reaction kinetics. When based on bond graph
models, the approach ensures thermodynamic compliance.

\S~\ref{sec:background} gives the equations required in the rest of
the paper.
\S~\ref{sec:reaction-kinetics} shows how conventional mass-action
kinetics, expressed in terms of forward $\Phi^f$ and reverse $\Phi^r$
reaction affinities, can be re-expressed in terms of reaction affinity
$\Phi = \Phi^f - \Phi^r$ and summed reaction affinity
$\sPhi = \Phi^f + \Phi^r$. This involves the hyperbolic functions
$\sinh$ and $\tanh$.
\S~\ref{sec:equilibria} shows how the equilibrium values of the  free
potentials can be deduced from the fixed potentials.
\S~\ref{sec:steady-states} contains the main results of the paper:
how the steady-state values of the free potentials can be deduced from
the fixed potentials for a given pathway flow.
\S~\ref{sec:exampl-biom-cycle} uses the biomolecular cycle of
\citet{Hil89} to illustrate the approach.
\S~\ref{sec:exampl-oxid-phosph} uses the Mitochondrial Electron Transport
Chain as a more detailed example.
\S~\ref{sec:conclusion} concludes the paper and suggest future research directions.

\section{Background}
\label{sec:background}
This section contains the background material required in the rest of
the paper. More information is to found in a companion
paper~\citep{Gaw17a}.

The Faraday-equivalent potential~\citep{Gaw17a} of a substance A is
given in terms of the potential $\phi_A^\std$ (with units of
Volts~\si{V}) when the amount of the substance is $x_A^\std\si{mol}$
as:
\begin{xalignat}{2}
  \phi_A &= \phi_A^\std + \phi_N \ln \frac{x_A}{x^\std_A} \label{eq:phi_A}&
\text{where } 
  \phi_N &= \frac{RT}{F} 
\end{xalignat}
$R = \SI{8.314}{JK^{-1}mol^{-1}}$ is the universal gas constant,
$T~\si{K}$ is the absolute temperature and
$F \approx \SI{96485}{C.mol^{-1}}$ is Faraday's constant.

\begin{enumerate}
\item For a given temperature, $\phi_N$ is a constant given by
  \eqref{eq:phi_A} and, at $T = \SI{300}{C}$ has a value of about
  $\SI{26}{mV}$.
\item Equation \eqref{eq:phi_A} has two parameters:
  \begin{enumerate}
  \item $\phi_A^\std$ the potential of the substance in the relevant
    compartment at nominal conditions and
  \item $x^\std_A$, the amount of the substance in the relevant
    compartment at the nominal
    conditions defining the nominal potential $\phi_A^\std$.
  \end{enumerate}
\item Given the two parameters $\phi_A^\std$ and $x^\std_A$, and the
  constant $\phi_N$, the amount $x_A$ may be deduced from $\phi_A$ by
  inverting Equation \eqref{eq:phi_A}:
  \begin{equation}
    x_A = x^\std_A \exp \frac{\phi_A - \phi_A^\std}{\phi_N} \label{eq:x_a}
  \end{equation}
\item The potential $\phi_A^\std$ of
  substance A at \emph{nominal} conditions can be derived from the
  potential $\phi_A^\Std$ at \emph{standard} conditions using the
  formula~\citep{AtkPau11}:
  \begin{equation}
    \phi_A^\std = \phi_A^\Std + \phi_N \ln \frac{x^\std_A}{x^\Std_A} 
  \end{equation}
Standard chermical potentials are listed for common substances by
\citet{AtkPau11}.
\item Equation \eqref{eq:phi_A} can be rewritten as:
  \begin{xalignat}{2}
    \phi_A &=  \phi_N \ln K_A x_A \label{eq:phi_A_K}&
    \text{where }
    K_A &= \frac{\exp \frac{\phi_A^\std}{\phi_N}}{x^\std_A}
  \end{xalignat}
\end{enumerate}

The forward $\Phi^f$ and reverse $\Phi^r$ reaction affinities can be
deduced from the Faraday-equivalent  potentials $\phi$ of the
substances using the forward $\Nf$ and reverse $\Nr$ stoichiometric
matrices as:
\begin{xalignat}{2}
  \Phi^f &= \Nf^T \phi \label{eq:Phi^f}&
  \Phi^r &= \Nr^T \phi 
\end{xalignat}
Furthermore, the net reaction affinity $\Phi$ is defined as
\begin{align}
  \Phi &= \Phi^f - \Phi^r \label{eq:Phi_def}\\
  \text{ and so }
  \Phi &= -\NN^T \phi \label{eq:Phi}
 \text{ where } 
  \NN = \Nr - \Nf
\end{align}
and $N$ is the system \emph{stoichiometric matrix}.
$N$, $\Nf$ and $\Nr$ can be automatically deduced from the system bond
graph \citep{GawCra14}.

Mass-action reaction kinetics can be written as:
\begin{align}
  v &= \kappa \lb \vp - \vm \rb\label{eq:MA_0}\\
  \text{where }
  \vp &= \exp \frac{\Phi^f}{\VN} \label{eq:vp}\\
  \text{and }
  \vm &= \exp \frac{\Phi^r}{\VN} \label{eq:vm}
\end{align}
where $\kappa$ is a constant and $\Phi^f$ and $\Phi^r$ are the forward
and reverse reaction affinities. 

From equations \eqref{eq:MA_0}~--~\eqref{eq:vm}, zero reaction flow $v=0$ implies that
\begin{xalignat}{2}
  \vp &= \vm & 
\text{and }
  \Phi &= \Phi^f - \Phi^r =0 \label{eq:Phi_0}
\end{xalignat}

As discussed in the literature, the system ode can be written as:
\begin{equation}
  \dX = N V
\end{equation}
where $\dX$ is the rate of change of the vector of chemical species
and electrical charge, $N$ is the system stoichiometric matrix
and
$V$ is the vector of chemical and
electrical flows.  

The notion of a \emph{chemostat}~\citep{PolEsp14,GawCra16}, the
chemical analogue of a voltage source, is useful when considering the
steady-states of open biomolecular systems.
In particular, as discussed by \citet{GawCra16}, in the presence of
chemostats:
\begin{equation}\label{eq:dX_cs}
  \dX = \Ncd V
\end{equation}
where $\Ncd$ is $N$ with the rows corresponding to the chemostats set
to zero.  In either case, the reaction potentials $\Phi$ are given by
Equation \eqref{eq:Phi}.

The left null space matrix $G$ of $\Ncd$ has the property that:
\begin{align}
  G\Ncd &= 0\\
  \text{hence }
  G\dX &= G\Ncd V = 0
\end{align}
The $n_{cm}$ rows of $G$ thus define the $n_{cm}$ \emph{conserved
  moieties} given by:
\begin{align}\label{eq:cm}
  GX = GX_0
\end{align}
where $X_0$ is the initial state.

In this paper, the \emph{order}  of the $i$th conserved moiety is
defined as the the number of non zero entries in the $i$th row of $G$
minus $1$. Thus zero-order moieties correspond to chemostats and first
order moieties to two substances whose net amount is constant.


\section{Reaction Kinetics}
\label{sec:reaction-kinetics}
%
The mass-action kinetics of eqns.~\eqref{eq:MA_0}~--~\eqref{eq:vm} are expressed in
terms of the forward and reverse reaction affinities $\Phi^f$ and
$\Phi^r$. One property of eqns.~\eqref{eq:MA_0}~--~\eqref{eq:vm} is that flow $v$ is
zero if $\Phi^f=\Phi^r$ or $\Phi=\Phi^f-\Phi^r=0$ where $\Phi$ is the
(net) reaction affinity.

To obtain further insight into mass-action reaction kinetics it is
therefore convenient to rewrite eqn.~\eqref{eq:MA_0} explicitly in terms of the
reaction affinity $\Phi$ and \emph{summed reaction affinity} $\sPhi$ where:
\begin{xalignat}{2}
  \Phi &= \Phi^f - \Phi^r&
  \text{and }
  \sPhi &= \Phi^f + \Phi^r\label{eq:sPhi}
\end{xalignat}
These definitions imply that 
\begin{xalignat}{2}
  \Phi^f &= \frac{\sPhi + \Phi}{2}&
  \text{and }
  \Phi^r &= \frac{\sPhi - \Phi}{2}
\end{xalignat}
and hence that
\begin{xalignat}{3}
  \vp &= \skappa \exp \frac{\Phi}{2\VN} &
  \text{and }
  \vm &=  \skappa \exp \frac{-\Phi}{2\VN}\\
  \text{where }
  \skappa &= \exp \frac{\sPhi}{2\VN}\label{eq:skappa}
\end{xalignat}
Introducing the difference $\vv$ and sum $\svv$ of $\vp$ \eqref{eq:vp}
and $\vm$ \eqref{eq:vm}:
\begin{align}
  \vv &= \vp - \vm = 2\skappa \sinh \frac{\Phi}{2\VN}\label{eq:v_0}\\
  \svv &= \vp + \vm = 2\skappa \cosh \frac{\Phi}{2\VN}\label{eq:sv_0}
\end{align}
where $\sinh$ and $\cosh$ are the hyperbolic $\sin$ and $\cos$
functions respectively.

Using these equations, the mass-action kinetics of eqn.~\eqref{eq:MA_0} can be
rewritten in two alternative forms:
\begin{xalignat}{1}
  v &= 2\kappa \skappa \sinh \frac{\Phi}{2\VN} \label{eq:MA_sinh}\\
  v &= \kappa \svv \tanh \frac{\Phi}{2\VN}   \label{eq:MA_tanh}
\\
  \text{where } \skappa &= \exp \frac{\Phi^f + \Phi^r}{2\VN}\label{eq:skappa_1}\\
  \text{and } \svv &= \exp \frac{\Phi^f}{\VN} + \exp \frac{\Phi^r}{\VN}\label{eq:svv}
\end{xalignat}
$\tanh$ is the hyperbolic $\tan$ function, 
\eqref{eq:skappa_1} follows from \eqref{eq:sPhi}\&\eqref{eq:skappa}
and \eqref{eq:svv} follows from Equations
\eqref{eq:vp}\&\eqref{eq:vm}.
Equation \eqref{eq:MA_sinh} is explicitly in terms of the reaction
affinity $\Phi$ \eqref{eq:Phi} and, via $\skappa$ \eqref{eq:skappa},
$\sPhi$ \eqref{eq:sPhi} and via \eqref{eq:Phi^f}, the potentials
$\phi$.
Equation \eqref{eq:MA_tanh} is also explicitly in terms of  the reaction
affinity $\Phi$ \eqref{eq:Phi} but, unlike Equation
\eqref{eq:MA_sinh}, depends on $\svv$ rather than $\skappa$.

Define $\skappaeq$ and $\svveq$ as the equilibrium values of
$\skappa$ and $\svv$ respectively.
At equilibrium, $\Phi^f=\Phi^r$ and thus, from Equations
\eqref{eq:skappa_1} and \eqref{eq:svv} it follows that
\begin{equation}
  \svveq = 2\skappaeq 
\end{equation}

Equations \eqref{eq:MA_sinh} and
\eqref{eq:MA_tanh} can be rewritten as:
\begin{xalignat}{2}
  v &=  2\kappabar \rho_s\sinh \frac{\Phi}{2\VN} &
  \text{where } 
   \rho_s &= \frac{\skappa}{\skappaeq}\label{eq:MA_sinh_0}\\
  v &= 2\kappabar \rho_t\tanh \frac{\Phi}{2\VN} &
  \text{where } 
  \rho_t &= \frac{\svv}{\svveq}
  \label{eq:MA_tanh_0}
\end{xalignat}
and $\kappabar = \kappa \skappaeq = \kappa \frac{\svveq}{2}$.
Near equilibrium, $\Phi \approx 0$ and these two equations can each be
approximated by functions of $\Phi$ only as:
\begin{align}
  v &\approx  2\kappabar \sinh \frac{\Phi}{2\VN}\label{eq:MA_sinh_a}\\
  v &\approx  2\kappabar \tanh \frac{\Phi}{2\VN}\label{eq:MA_tanh_a}
\end{align}
They may be further approximated by the linear equation:
\begin{align}
  v &\approx  \kappabar \frac{\Phi}{\VN} 
\end{align}

\subsection{Example: A Basic Electrogenic Network}
\label{sec:exampl-basic-electr}
\begin{figure}[htbp]
  \centering
  \Fig{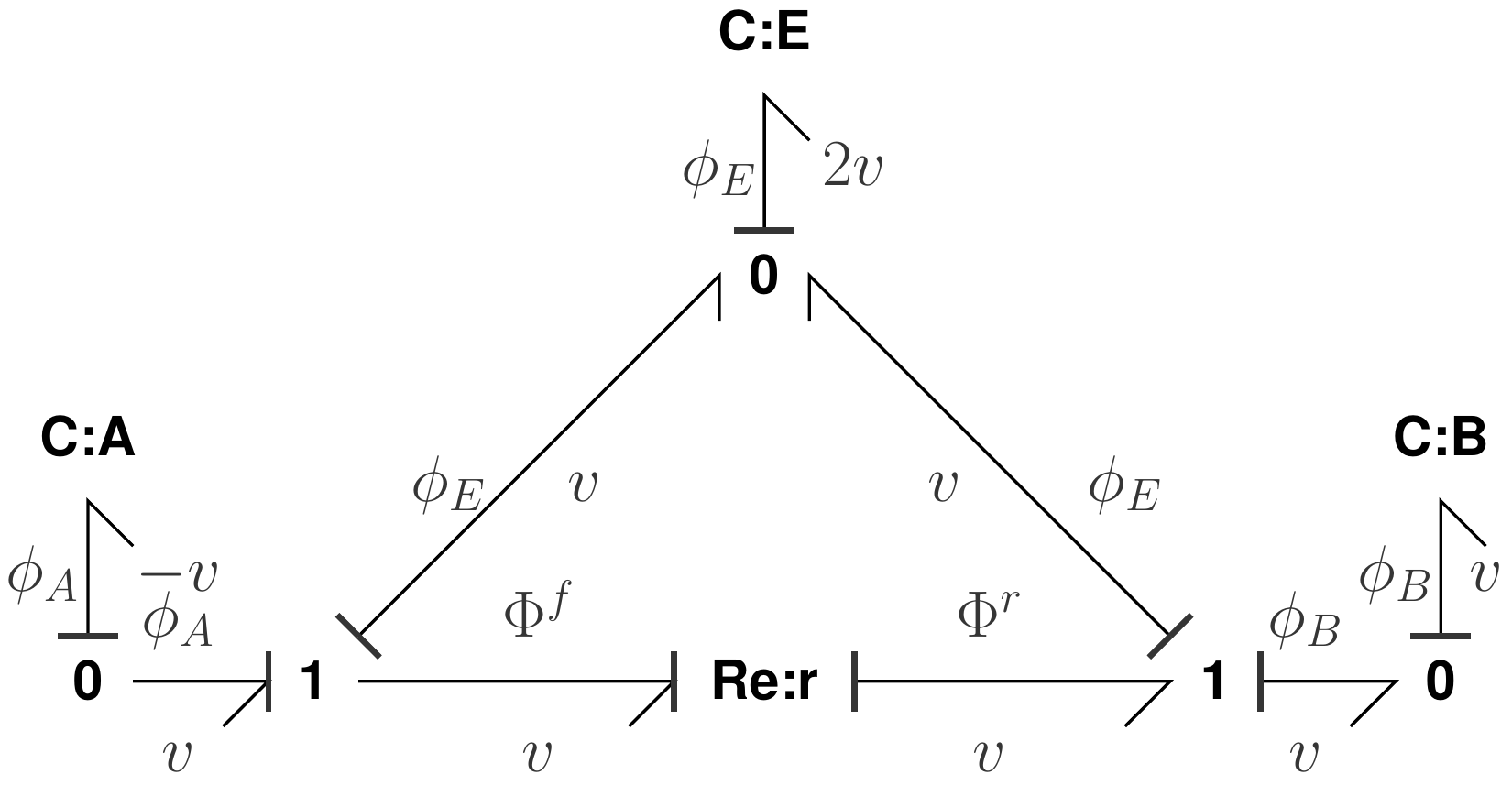}{0.6}
  \caption[A Basic Electrogenic Network]{A Basic Electrogenic
    Network. The system bond graph has two \C components: \BC{A} and
    \BC{B} representing the chemical species A and B respectively and
    a third \C component \BC{E} representing electrical trans-membrane
  charge. The \Re component represents the reaction and the \zero and
  \one junction connect components appropriately. The bonds have been
  annotated to show the corresponding Faraday-equivalent potential
  (effort variable) and Faraday-equivalent current (flow
  variable). Thus the bond impinging on the \BRe{r} component has the
  effort variable $\Phi^f=\phi_A-\phi_E$ and flow variable $v$.
  The constitutive relations for the \C components are given by
  \eqref{eq:phi_A} and for the \Re component by
  \eqref{eq:MA_0}~--~\eqref{eq:vm}.
}
\label{fig:ABE_BG}
\end{figure}
Figure \ref{fig:ABE_BG} gives the bond graph  of a basic
electrogenic network where the trans-membrane reaction:
\begin{equation}
  A \reacu{r} B
\end{equation}
accumulates charge on the capacitor \BC{E}. 
For illustration, it is assumed that:
\begin{equation}
  \phi_A^\std = \phi_B^\std = 0
\end{equation}

The state vector is chosen as
\begin{align}
  X &=
      \begin{pmatrix}
        x_A&x_B&x_E
      \end{pmatrix}^T
\end{align}
where $x_A$, $x_B$ and $x_E$ are the amounts of $A$, $B$ and charge
respectively.  The stoichiometric matrices $N$, $\Nf$ and $\Nr$ are
automatically derived from the bond graph of Fig. \ref{fig:ABE_BG} and
are given by:
\begin{xalignat}{3}
  N &=
  \begin{pmatrix}
    -1\\1\\2
  \end{pmatrix}&
  \Nf &=
  \begin{pmatrix}
    1\\0\\-1
  \end{pmatrix}&
  \Nr &=
  \begin{pmatrix}
    0\\1\\1
  \end{pmatrix}
\end{xalignat}
Hence the corresponding reaction affinities are given by Equations
\eqref{eq:Phi^f}~--~\eqref{eq:Phi}
as
\begin{xalignat}{2}
  \Phi^f &= \phi_A - \phi_E&
  \Phi^r &= \phi_B + \phi_E\label{eq:Phir_example}
\end{xalignat}
and the corresponding difference and summed reaction affinities are:
\begin{align}
  \Phi &= \phi_A - \phi_B - 2\phi_E&
  \sPhi &= \phi_A + \phi_B \label{eq:Phi_example}
\end{align}

\begin{figure}[htbp]
  \centering
  \SubFig{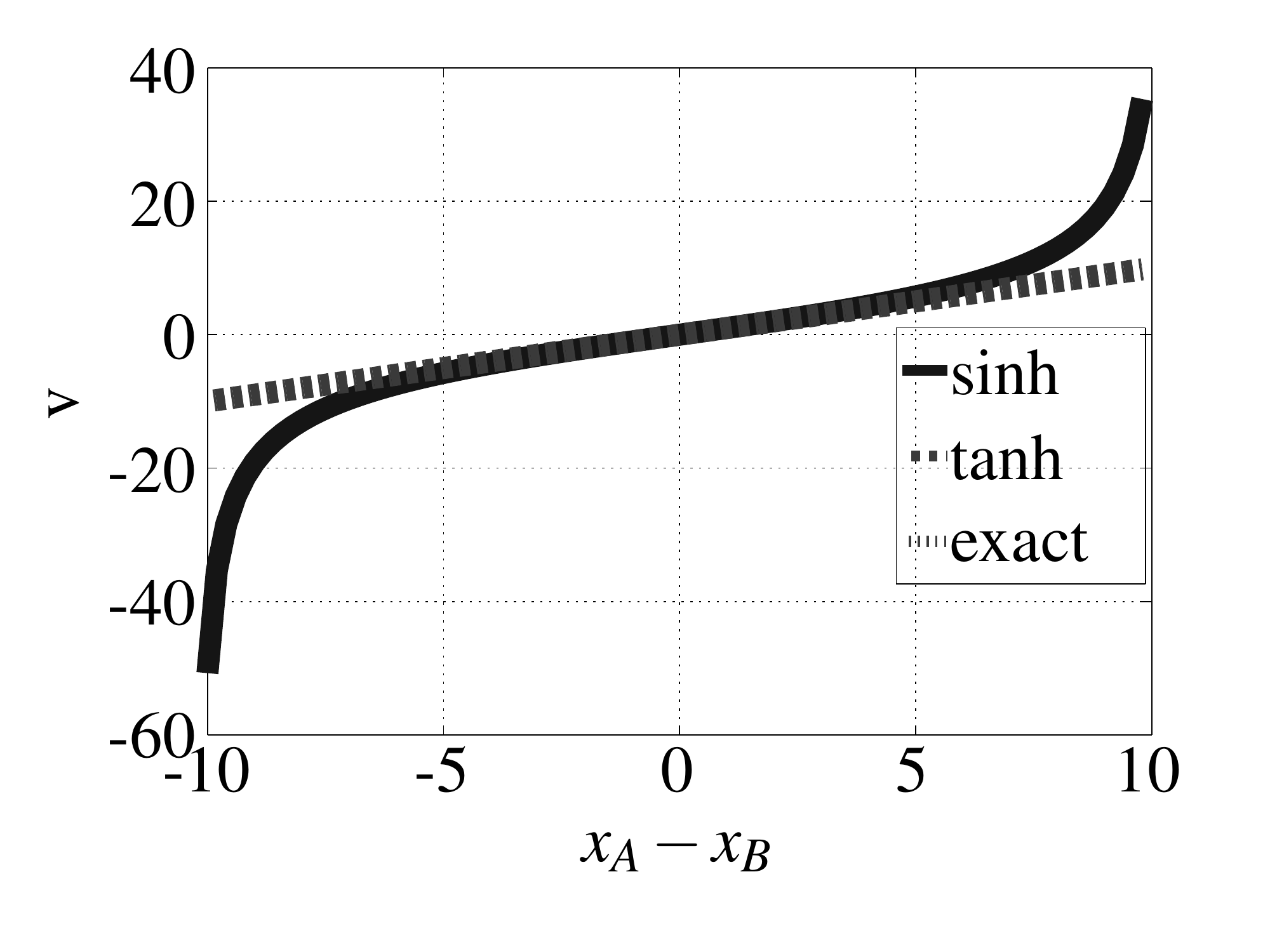}{$x_A+x_B=10.2$}{\FigWidth}
  \SubFig{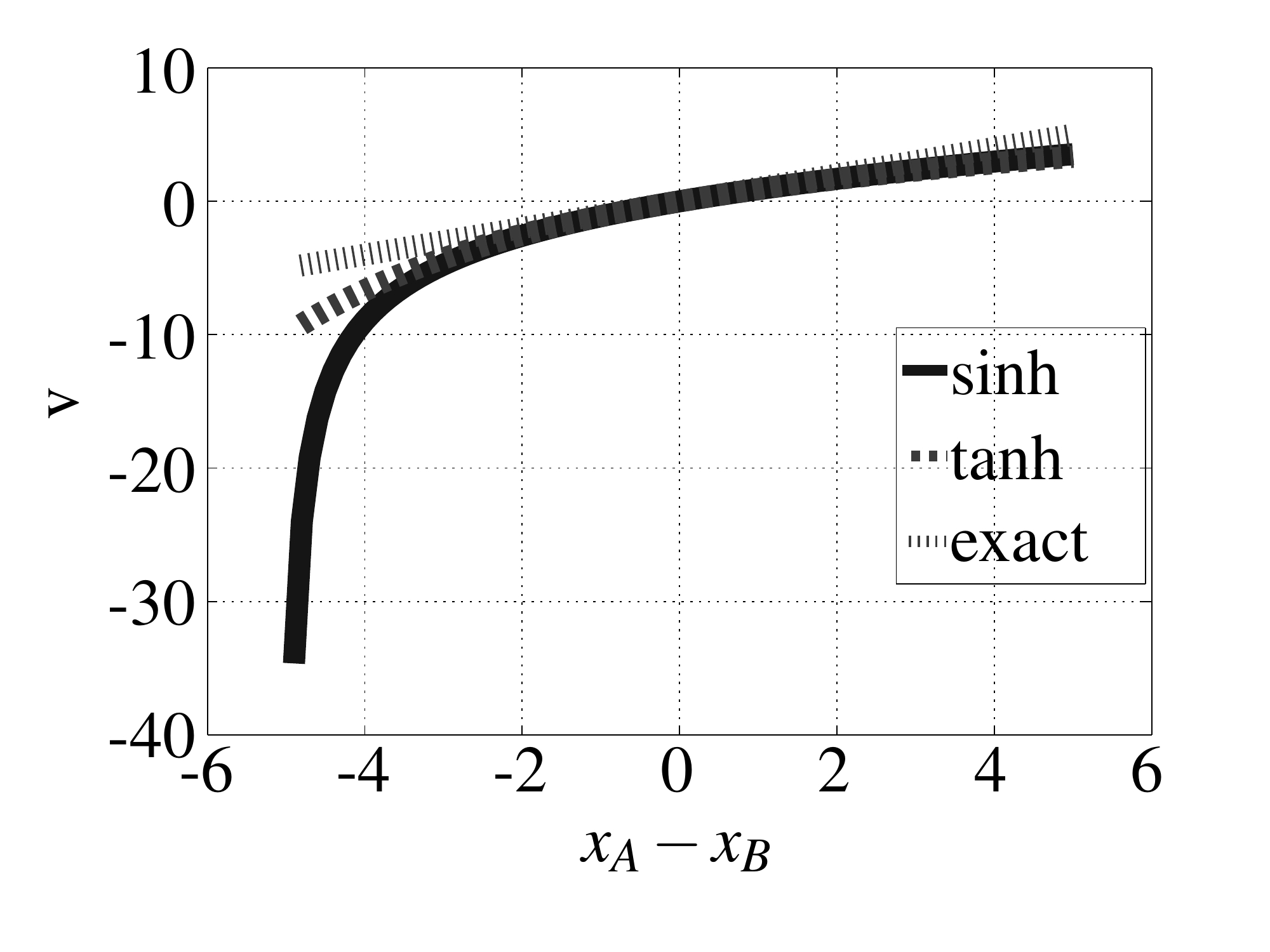}{$x_B=5$}{\FigWidth}
  \caption[Non-electrogenic case.]{Non-electrogenic case. 
    Flow $v$ computed exactly using Octave \citep{EatBatHau15} and using the $\sinh$ and  $\tanh$
    approximations.
    (a)  $x_A+x_B=10.2$. The $\tanh$ approximation is exact in this case.
    %
    %
    (b)  $x_B=5$. Neither approximation is exact in this case.
  }
\label{fig:RK-non-electrogenic}
\end{figure}
Consider first the non-electrogenic case where $\phi_E=0$. Then using
the formula \eqref{eq:phi_A} and eqn. \eqref{eq:sv_0} for chemical
potential:
\begin{align}
  \svv &= \exp \frac{\Phi^f}{\VN} + \exp \frac{\Phi^r}{\VN}
       = \exp \frac{\phi_A}{\VN} + \exp \frac{\phi_B}{\VN}\notag\\
       &= K_A x_A + K_B x_B
\end{align}
For the purposes of illustration, further assume that $K_A=K_B=1$ thus
\begin{equation}
  \svv = x_A + x_B
\end{equation}
As $A$ and $B$ form a conserved moiety,  $\svv$ is constant and thus
the $\tanh$ approximation \eqref{eq:MA_tanh_a} is exact.  Figure
\ref{subfig:CR_CM} shows the exact flow $v$ together with the
approximations \eqref{eq:MA_sinh_a} and \eqref{eq:MA_tanh_a} for this
case.

In contrast, Figure \ref{subfig:CR_noCM} shows the case where $x_b=0.5$
and $x_a$ varies. There is \emph{no} conserved moiety:
$x_A+x_B \ne constant$.  In this case neither approximation is exact,
though the $\tanh$ approximation is better than the $\sinh$
approximation.

\begin{figure}[htbp]
  \centering
  \Fig{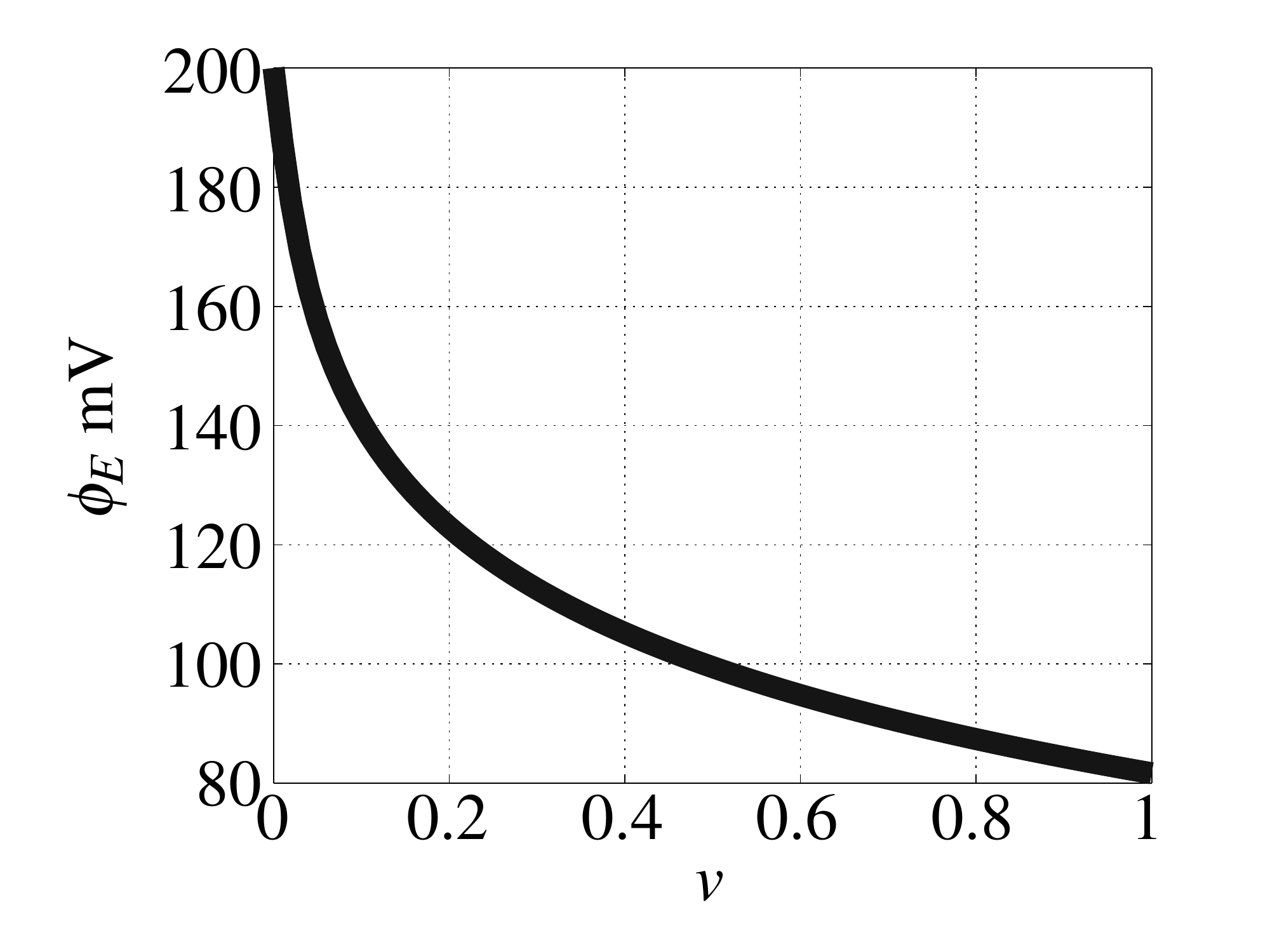}{\FigWidth}
  \caption{Electrogenic case with fixed concentrations. 
    Using Octave \citep{EatBatHau15}, the capacitor potential $\phi_E$
    plotted against reaction flow when $\phi_A=\SI{500}{mV}$ and
    $\phi_B=\SI{100}{mV}$. 
    The reaction constant \eqref{eq:MA_sinh_0} is chosen as 
    $\kappa_s = 0.01$.
    At $v=0$, $\phi_E = \frac{\phi_A - \phi_B}{2}=\SI{200}{mV}$; as
    $v$ increases, $\phi_E$ decreases as the inverse hyperbolic sin of
    \eqref{eq:phi_E_form}.}
\label{fig:RK-electrogenic}
\end{figure}
Now consider the electrogenic, but constant concentration, case where
$\phi_E$ varies but $\phi_A$ and $\phi_B$ are constant. It follows
from Equation \eqref{eq:Phi_example} that $\skappa$ is constant and
thus the $\sinh$ approximation \eqref{eq:MA_sinh_a} is exact.
Combining equations \eqref{eq:MA_sinh_a} and \eqref{eq:Phi_example},
it follows that:
\begin{align}\label{eq:phi_E_form}
  \phi_E &= \frac{\phi_A - \phi_B}{2} - \phi_N \asinh \frac{v}{2 \kappabar}
\end{align}
This is plotted in Figure \ref{fig:RK-electrogenic} for a particular
choice of parameters. This result is qualitatively similar to the
experimentally obtained results of \citet[Figure 2B]{BazBeaVin16} for
the Mitochondrial Electron Transport Chain. This is explored further
in \S~\ref{sec:exampl-oxid-phosph}.

\section{Equilibria}
\label{sec:equilibria}
In typical biomolecular systems, the reaction kinetics of
\S~\ref{sec:reaction-kinetics} are  embedded in large networks. This
section considers the equilibria of such networks.
In this paper, a system is said to be in \emph{equilibrium} if efforts
(eg chemical potentials and voltages) are such that all flows are
zero,
\begin{equation}\label{eq:equilibrium}
  \VV = 0
\end{equation}
which implies from the arguments leading to equation \eqref{eq:Phi_0}
that the reaction affinities $\Phi=0$.

As discussed in \S~\ref{sec:background}, the affinities $\Phi$ are
given by  equation \eqref{eq:Phi} ($\Phi = -N^T \phi$).
Equation \eqref{eq:Phi} is linear and therefore finding the potentials
$\phi$ to give zero affinities $\Phi=0$ is a linear problem.  However,
the $n_V \times n_X$ matrix $N^T$ is usually not full rank as usually
$n_V < n_X$ and therefore \eqref{eq:Phi} cannot be used directly to
deduce the potentials $\phi$ from reaction affinities $\Phi$.

However, the conserved moieties of Equation \eqref{eq:cm} give a
further set of constraints on the potentials $\phi$. In particular,
using Equation \eqref{eq:x_a}, Equation \eqref{eq:cm} becomes:
\begin{align}
   G \lb X^\std \hp  \exp  \frac{\phi -  \phi^\std}{\phi_N} \rb &= GX_0
\label{eq:GX_phi}
\end{align}
where $\hp$ denotes elementwise multiplication.
Equation \eqref{eq:GX_phi} is \emph{nonlinear} in $\phi$.

\subsection{Conserved Moieties with order zero}
In the special case that all conserved moieties are order $0$,
Equation \eqref{eq:GX_phi} can be rewritten as the linear equation.
\begin{align}
  G \phi &= \phi^{cm}\label{eq:Gphi}\\
  \text{where }
  \phi^{cm} & = G \phi_0
  \text{ and }
  \phi_0 = \phi^\std + \phi_N \ln \frac{X_0}{X^\std}
\end{align}
In this case Equations \eqref{eq:Phi} and  \eqref{eq:Gphi} can be
combined as:
\begin{equation}\label{eq:phi_cm_0}
  \begin{pmatrix}
    -N^T\\G
  \end{pmatrix}
  \phi =
  \begin{pmatrix}
    \Phi\\\phi^{cm}
  \end{pmatrix}
\end{equation}
Once again, equilibria are defined by setting $\Phi=0$ and the
solution of Equation \eqref{eq:phi_cm_0} is a linear problem. 


\subsection{Conserved Moieties with order greater than zero}
\label{sec:conserved-moieties}
The order-zero case gives the linear set of equations
\eqref{eq:phi_cm_0}. This section shows how the non-linear equations
arising when the conserved moiety is not zero-order may be solved.

Define $G_{cm}$ as the matrix containing the non-zero-order rows of $G$.
\begin{enumerate}
\item The non-zero-order moieties are converted to order-zero by deleting
  all but one element of the rows of $G_{cm}$.
\item A $\phicm$ for the new chemostats is chosen and the linear problem
  \eqref{eq:phi_cm_0} solved. 
\item Compute $\Gcm X$ and evaluate discrepancy with respect to $X^{cm}$.
\item Choose new $\phicm$ to reduce discrepancy and repeat from step 2.
\end{enumerate}
This is conveniently automated using \texttt{fsolve} in
Octave~\citep{EatBatHau15}.

\subsection{Example: A Basic Electrogenic Network
  (continued)}\label{sec:exampl-basic-electr-1}

\begin{figure}[htbp]
  \centering
  \SubFig{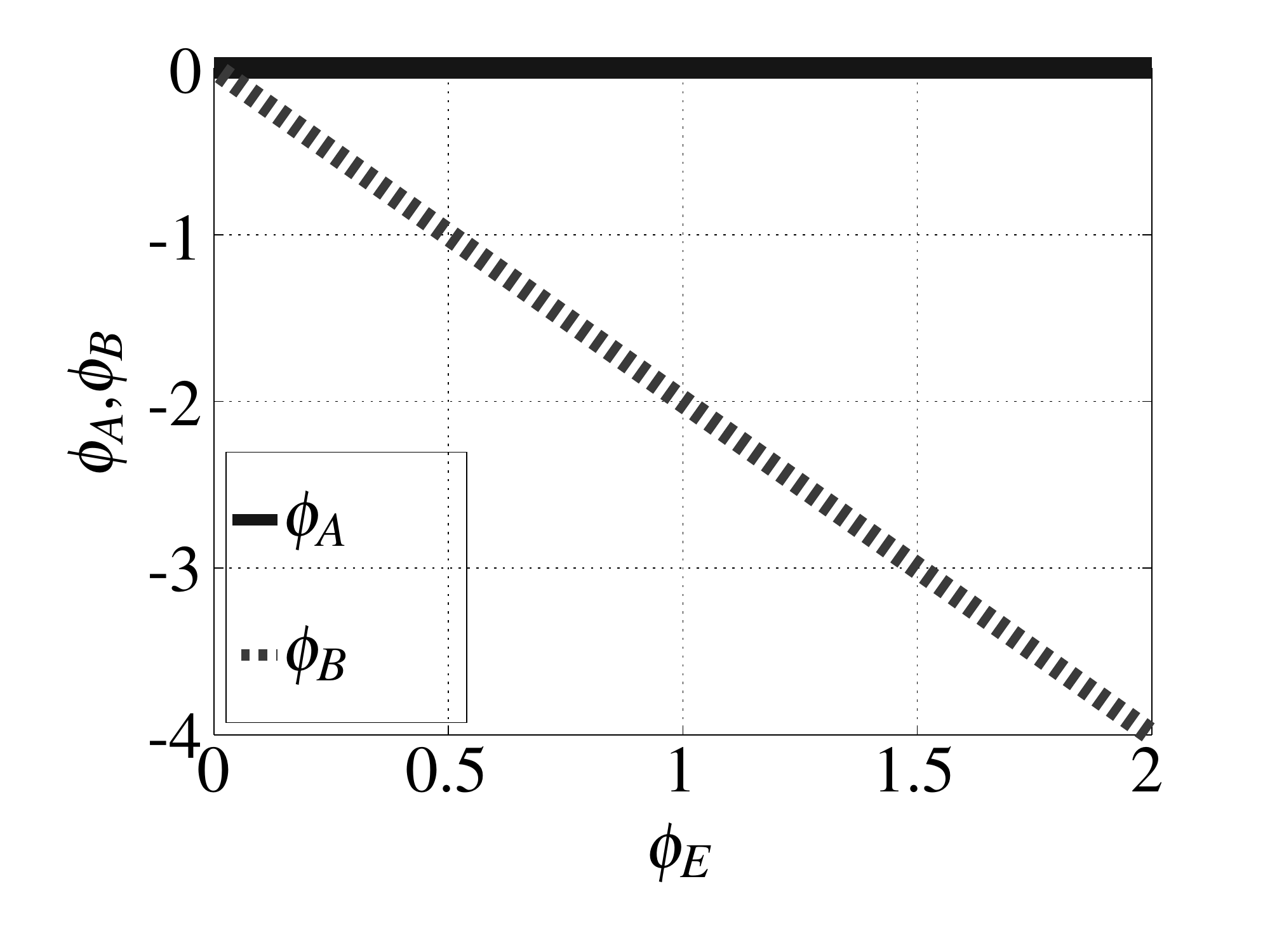}{$\phi_A$, $\phi_B$}{\FigWidth}
  \SubFig{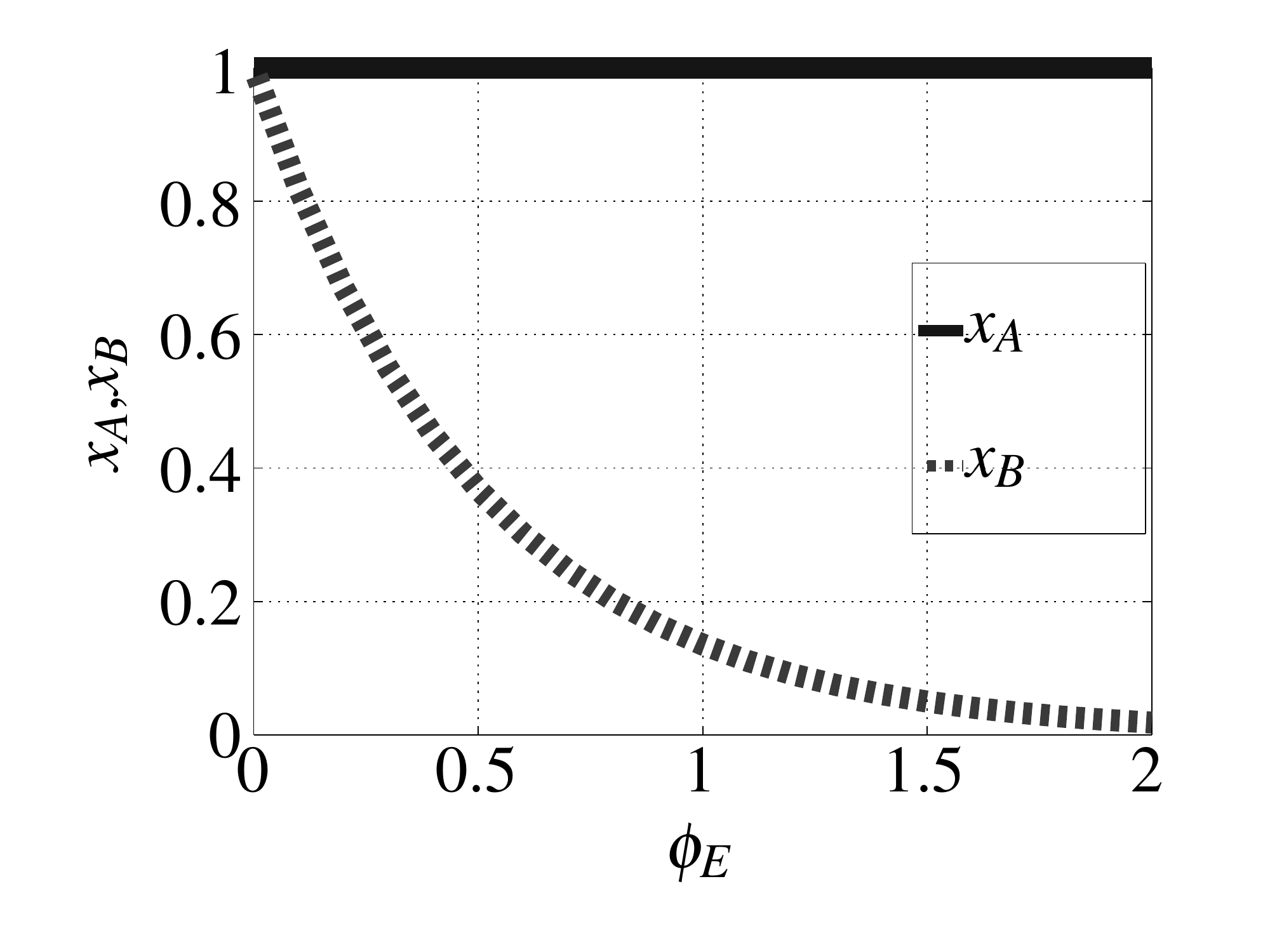}{$X$}{\FigWidth}
  \caption{Electrogenic system: Equilibria, order-zero moieties. A and
  E are chemostats; B is free to vary.
  (a) The potentials $\phi_A$ and $\phi_B$ when B is a chemostat with
  $\phi_B=0$ and the electrical potential $\phi_E$ varies.
  (b) The corresponding states $x_A$ and $x_B$.
}
\label{fig:nocm}
\end{figure}
Consider first the case where both A and E are chemostats. Then $\Ncd$
\eqref{eq:dX_cs} is given by $N$ with the first and third rows
deleted:
\begin{align}
 \Ncd &=
  \begin{pmatrix}
    0&1&0
  \end{pmatrix}^T
\end{align}
One choice of $G$ is
\begin{align}
  G &=
  \begin{pmatrix}
    1&0&0\\
    0&0&1
  \end{pmatrix}^T
\end{align}
As there is only one non-zero entry in each row of $G$,  the linear
equation \eqref{eq:phi_cm_0} can be used with:
\begin{xalignat}{2}
  \begin{pmatrix}
    -N^T\\G
  \end{pmatrix} &=
  \begin{pmatrix}
    -1&1&2\\
    1&0&0\\
    0&0&1
  \end{pmatrix};&
  \begin{pmatrix}
    \Phi\\\phi^{cm} 
  \end{pmatrix} &=
                  \begin{pmatrix}
                  0\\ \phi_A \\ \phi_E  
                  \end{pmatrix}
\end{xalignat}
Figure \ref{fig:nocm} shows the solution of equation
\eqref{eq:phi_cm_0} for $\phi_A$, $\phi_B$ plotted against $\phi_E$
when the potential chemostat A is fixed at $\phi_A=0$ and $\phi_E$
varies from 0 to 2. In this case the solution is simply that
$\phi_B=-2\phi_E$ as can be readily obtained from Equation
\eqref{eq:Phi_example} by setting $\Phi=\phi_A=0$.

\begin{figure}[htbp]
  \centering
  \SubFig{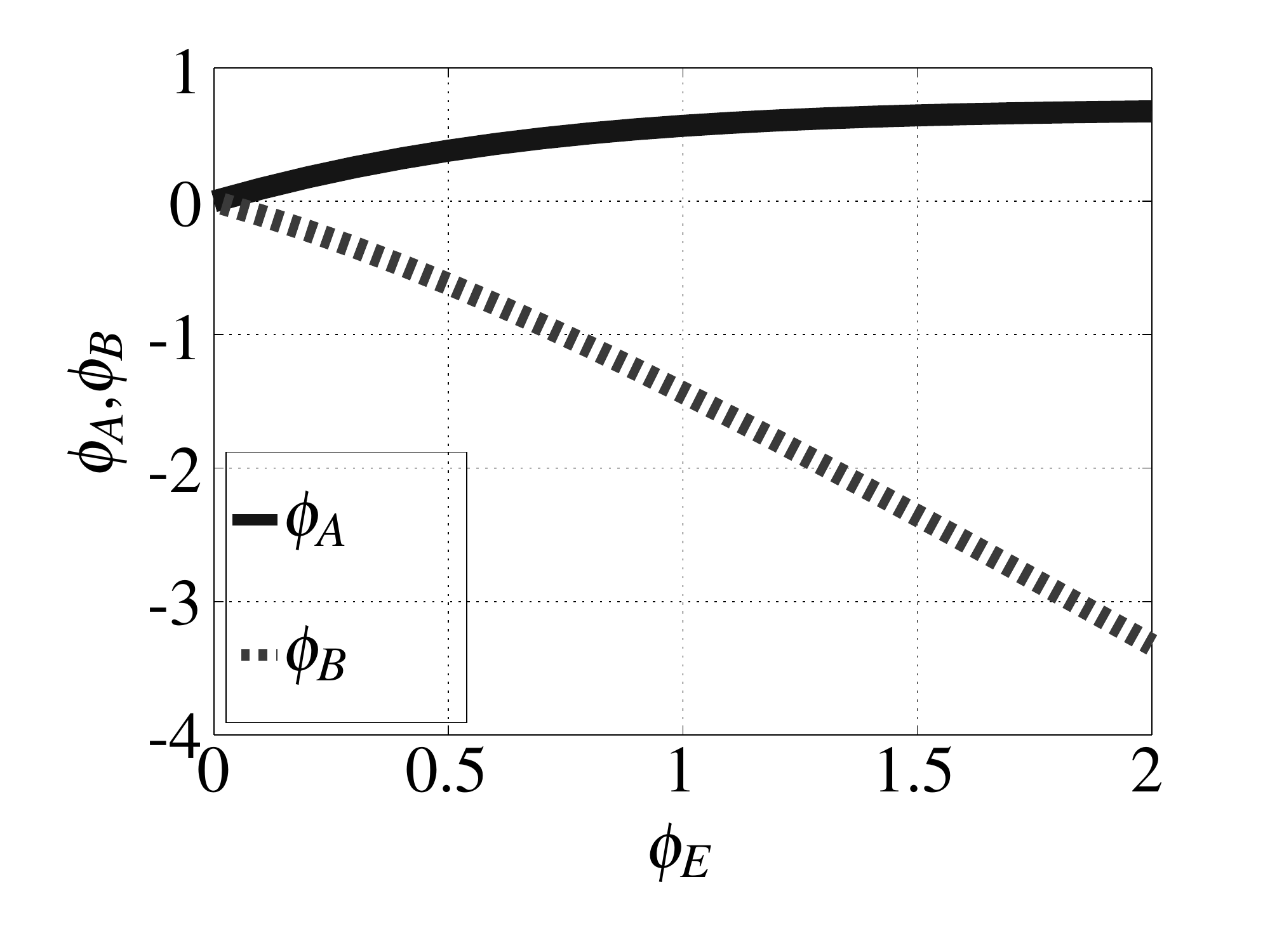}{$\phi_A$, $\phi_B$}{\FigWidth}
  \SubFig{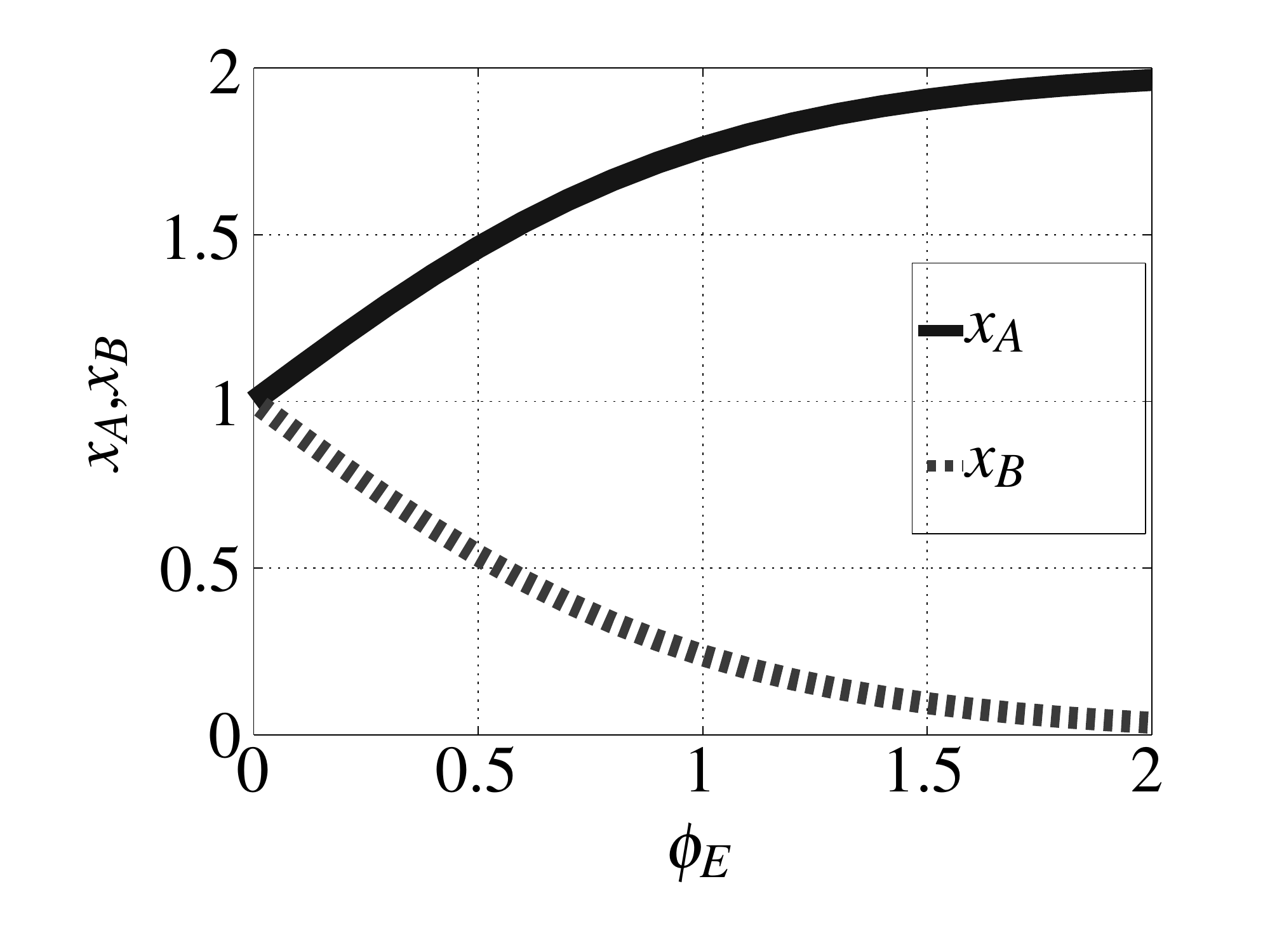}{$X$}{\FigWidth}
  \caption{Electrogenic system: Equilibria, order-one moieties. E is a
  chemostat, A \& B form a conserved moiety.
  (a) The potentials $\phi_A$ and $\phi_B$ when A and B form a
  conserved moiety with $x_A+x_B=2$ and the electrical potential $\phi_E$ varies.
  (b) The corresponding states $x_A$ and $x_B$. Note that $x_A+x_B=2$
  for all flows.}
\label{fig:cm}
\end{figure}
Consider secondly the case where only E is a chemostat. Then $\Ncd$
\eqref{eq:dX_cs} is given by $N$ with the third row
deleted:
\begin{align}
 \Ncd &=
  \begin{pmatrix}
    -1&1&0
  \end{pmatrix}^T
\end{align}
One choice of $G$ is
\begin{align}
  G &=
  \begin{pmatrix}
    1&1&0\\
    0&0&1
  \end{pmatrix}^T
\end{align}
As there are two non-zero entries in the first row of $G$, there is a
order-one conserved moiety and so the method of
\S~\ref{sec:conserved-moieties} must be used.
In this case
\begin{equation}
  \Gcm =
  \begin{pmatrix}
    1&1&0
  \end{pmatrix}
\end{equation}
Deleting the second element of the first row of $G$ gives the same
linear equation as for the two-chemostat case; the difference is that
$\phi_A$ must be chosen to satisfy the conserved moiety
$x_A+x_B=x_{AB}$. In this example, choose $x_{AB}=2$ and thus
$\phicm=\phi_N \ln 2$.  Figure \ref{fig:cm} shown the results of the procedure of
\S~\ref{sec:conserved-moieties} where $\phi_A$, $\phi_B$ plotted
against $\phi_E$ when the potential chemostat A is fixed at $\phi_A=0$
and $\phi_E$ varies from 0 to 2.

\section{Steady-states}
\label{sec:steady-states}
In this paper, a system is said to be in \emph{steady-state} if efforts
are such that all system states are constant. Thus in the context of
stoichiometric models:
\begin{equation}\label{eq:ss}
  \dX = \Ncd \VV = 0
\end{equation}
where $\Ncd$ is the \emph{chemodynamic} stoichiometric matrix
\citep{GawCra16}.  In the special case of equilibrium
\eqref{eq:equilibrium}, $\VV = 0$ and Equation \eqref{eq:ss} is
automatically satisfied.

Pathway analysis determines the set  of \emph{non-zero} flows such
that Equation \eqref{eq:ss} is satisfied. In particular such analysis
determines the pathway matrix $K_p$ such that flows $\VV$ given by 
\begin{align}
  \VV &= K_p v_p
\end{align}
satisfy Equation \eqref{eq:ss} for any choice of $v_p$.





\citet{GawCra17} present an energy-based approach to pathway
analysis. This forms the basis of the method discussed in this
section. In particular, the starting point for the method proposed
here is to apply positive-pathway analysis \citep{GawCra17} to
the system where the chemostats have been chosen to give exactly one
pathway described by $K_p$.

The method is as follows:
\begin{enumerate}
\item Choose a vector of pathway flows $v_p$ and deduce all
  the relevant steady-state flows from:
  \begin{equation}
    \VV = K_p v_p
  \end{equation}
\item Choose a starting value for the potentials $\phi$ -- for example
  that corresponding to equilibrium -- for all flows.
\item Compute the $\skappa$ -- based on $\phi$ -- for all flows.
\item \label{item:start} Deduce the reaction affinities by inverting  the approximation \eqref{eq:MA_sinh_a}
  \begin{equation}
    \Phi = 2\phi_N \asinh \frac{\VV}{2\kappabar}
  \end{equation}
\item Using the approach of \S~\ref{sec:equilibria}, deduce the
  free-to-vary potentials from Equation \eqref{eq:phi_cm_0}:
\begin{equation}
  \NN_v \phi_v = \Phi-\NN_{cs} \phi_{cs}
\end{equation}
\item If $\phi$ has not converged, then return to step \ref{item:start}.
\end{enumerate}

Assuming convergence, this method yields the steady-state potentials
$\phi$ as well as the corresponding $\skappa$.

An alternative approach would use the $\tanh$ formulation
\eqref{eq:MA_tanh} and $\svv$ in place of the $\sinh$ formulation
\eqref{eq:MA_sinh} and $\skappa$.

\section{A Biomolecular Cycle}\label{sec:exampl-biom-cycle}
\begin{figure}[htbp]
  \centering
  \SubFig{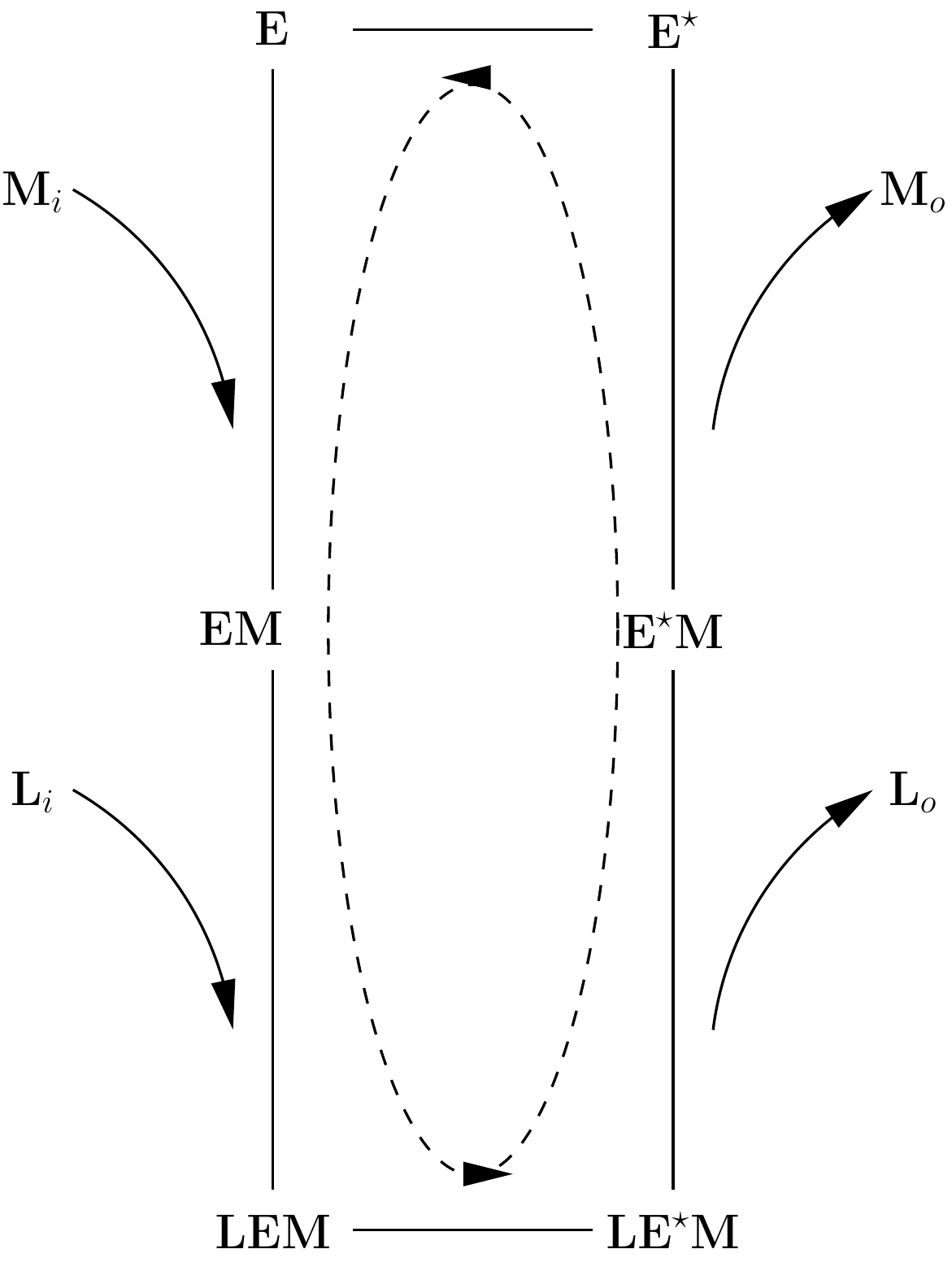}{Schematic}{0.5}
  \SubFig{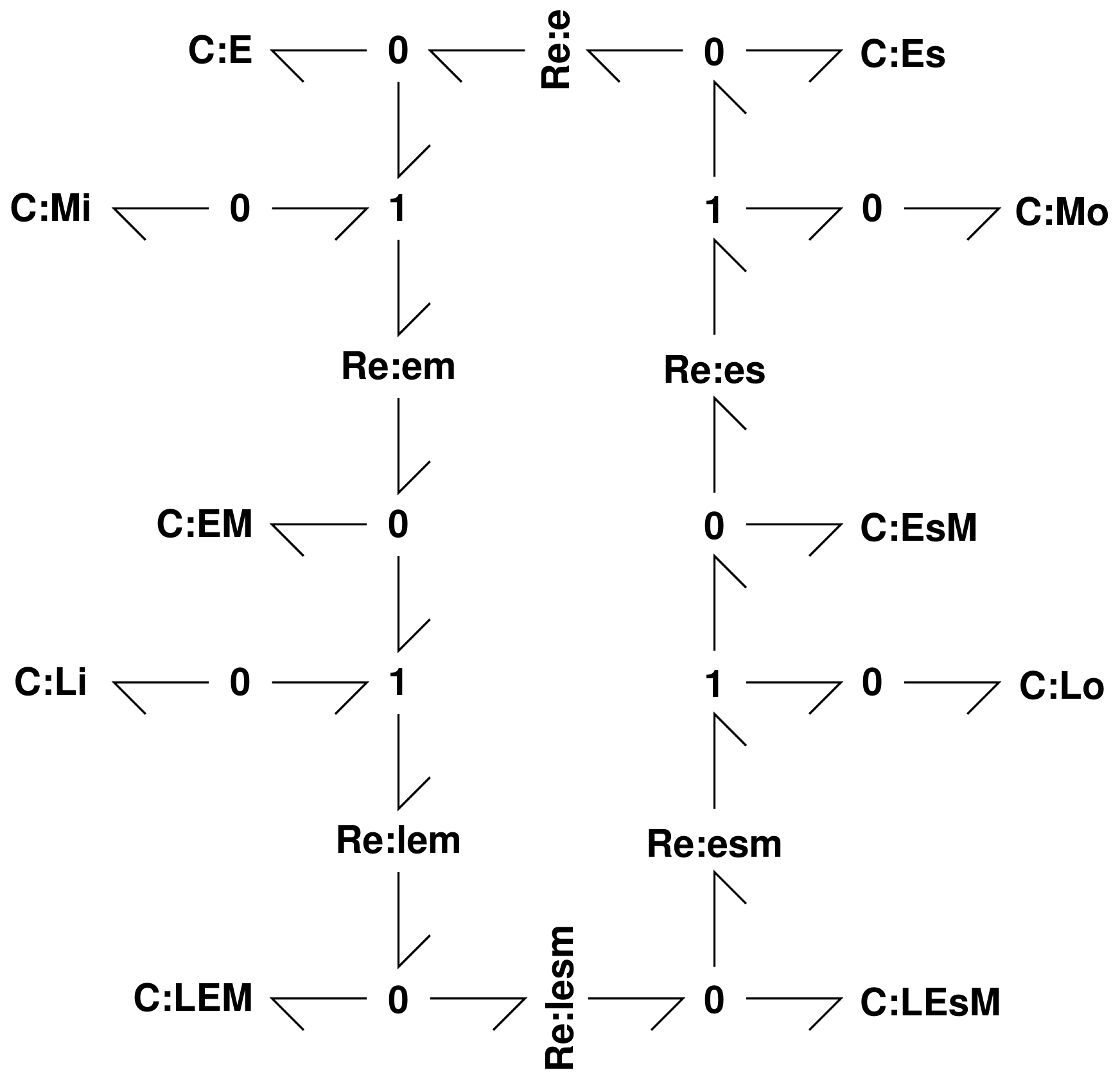}{Bond Graph}{\FigWidth}
  \caption[Example: A Biomolecular Cycle]{Example: A Biomolecular
    Cycle.
    (a) Schematic corresponding to \citep[Figure 1.1]{Hil89}.
    (b) Bond graph  corresponding to (a).
    (c) Bond graph  with the electrogenic capacitor \BC{V_m} corresponding
    to a trans-membrane voltage.
}\label{fig:Cycle}
\end{figure}
This example is based on that of \citet[\S~5]{GawCra17} which in turn
corresponds to the seminal monograph, ``Free energy transduction and
biomolecular cycle kinetics'' of \citet{Hil89}. As discussed by
\citet[\S~5]{GawCra17}, the biomolecular cycle of Figure 1.1 of
\citet{Hil89} may be represented as the bond graph of Figure
\ref{subfig:Hill_abg}. 
\begin{figure}[htbp]
  \centering
  \SubFig{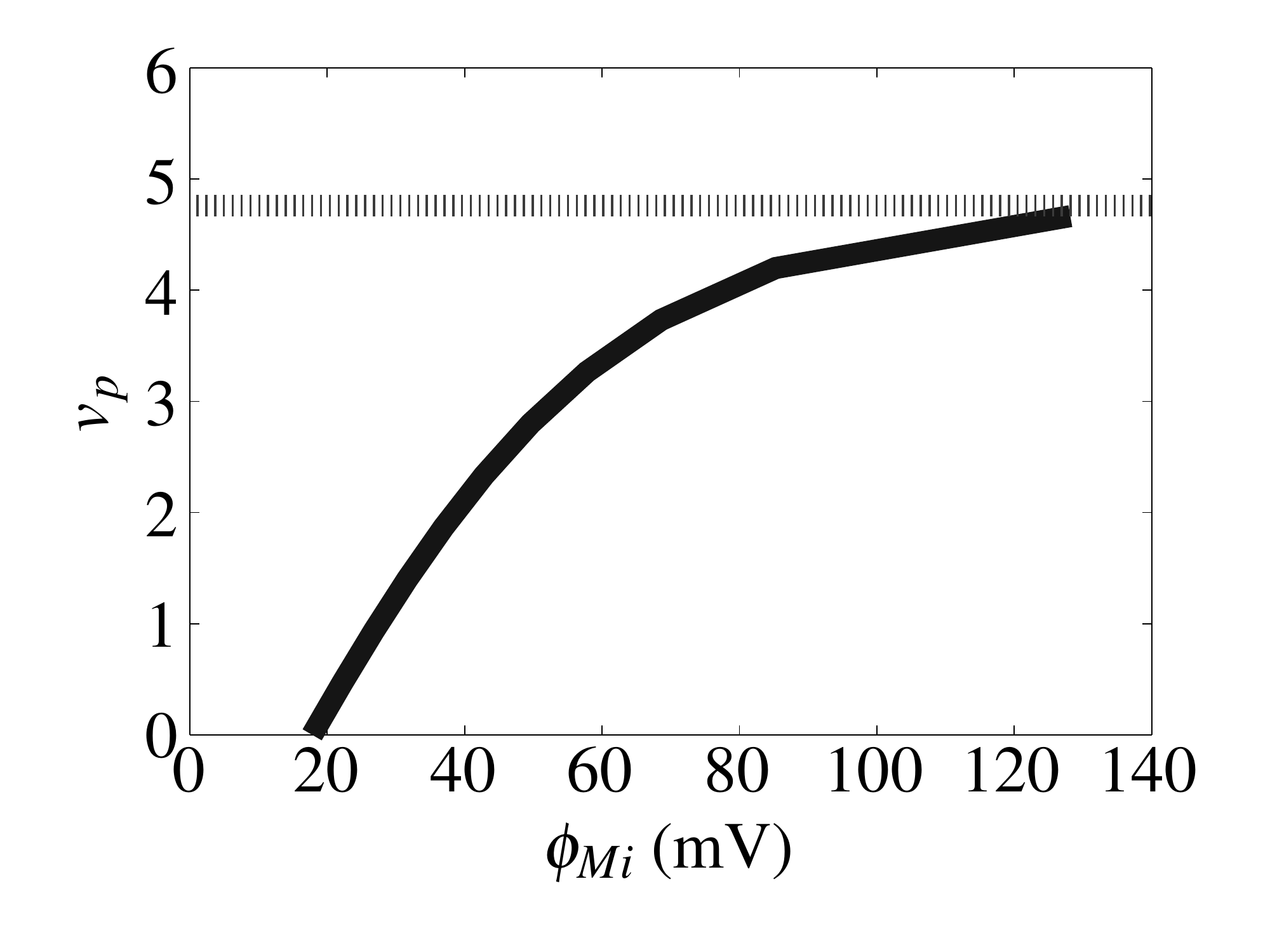}{Pathway flow $v_p$ and $\phi_{Mi}\si{~mV}$}{\FigWidth}
  \SubFig{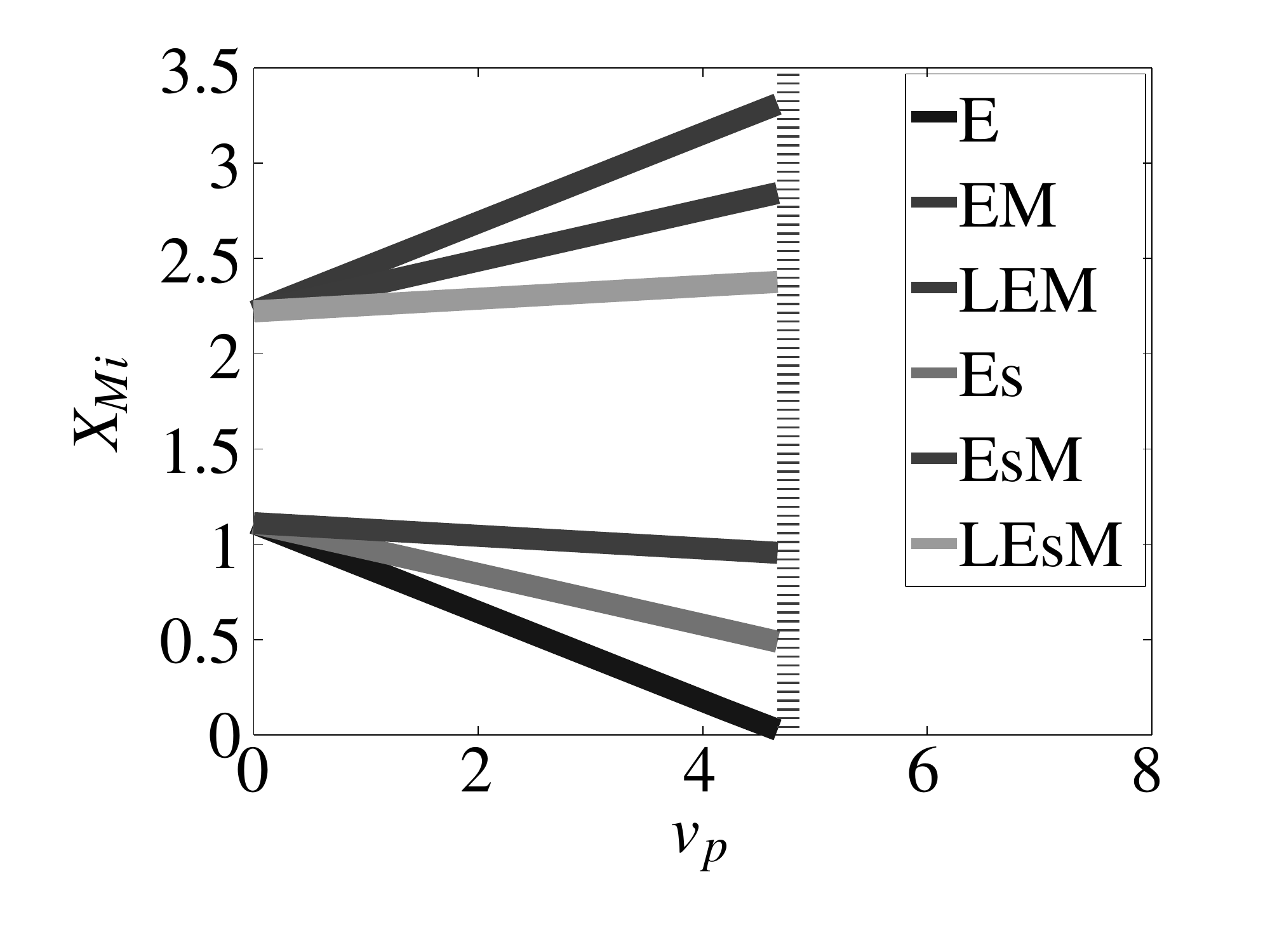}{Conserved moiety}{0.45}
  \SubFig{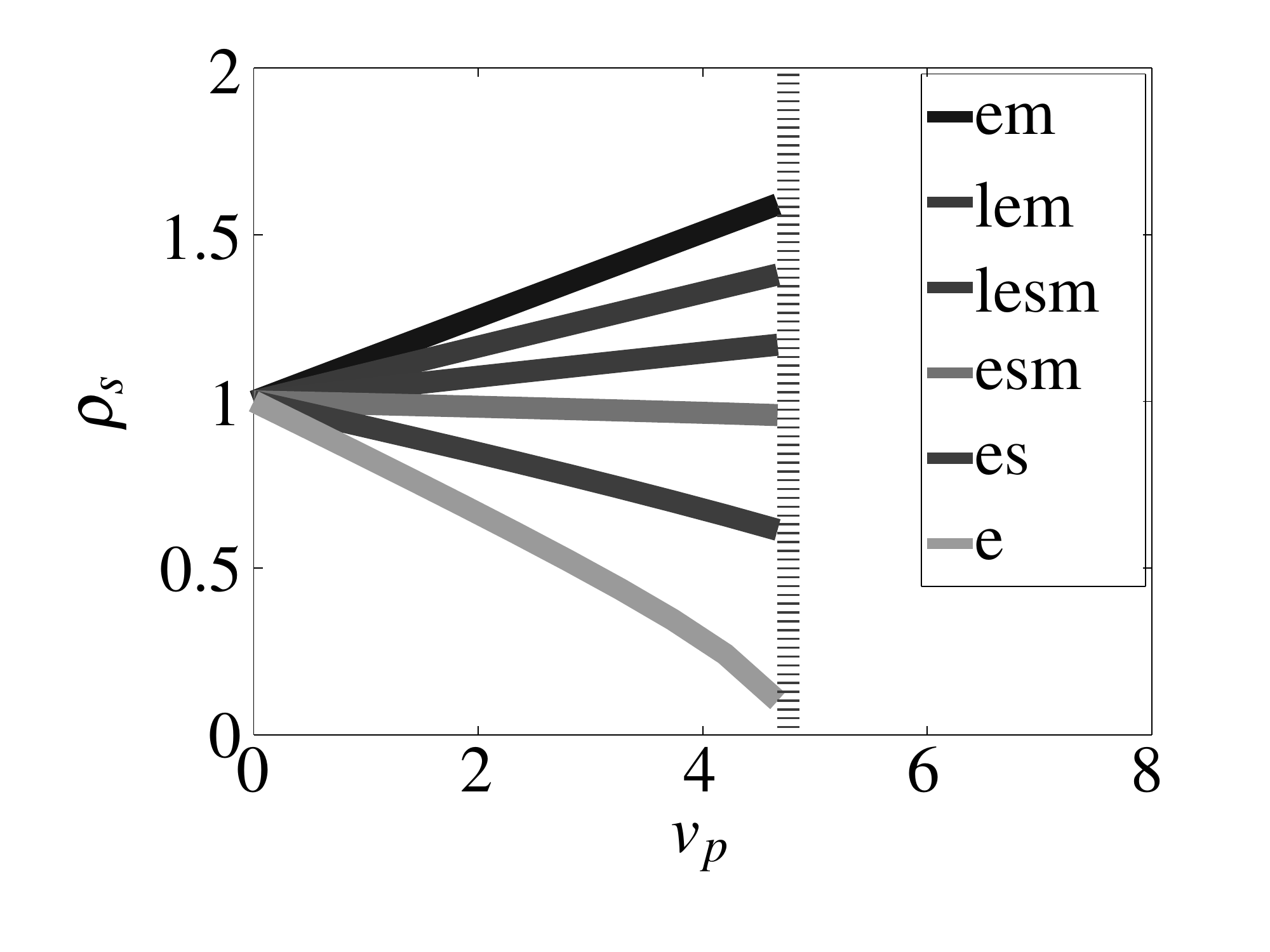}{$\rho_s$}{0.45}
  \caption[Example: A Biomolecular Cycle.]{Example: A
    Biomolecular Cycle.
    (a) The flow $v_P$ is plotted against the variable potential
    $\phi_{Mi}\si{~mV}$. The maximum flow $v_{max}$ is marked as a
    horizontal line.
    %
    %
    (b) The six states corresponding to the conserved moiety are
    plotted against the pathway flow $v_P$. The maximum flow $v_{max}$
    is marked by a vertical line an corresponds to the flow where a
    state ($x_E$ in this case) becomes zero.
    (c) The values of $\rho_s$ \eqref{eq:MA_sinh} for the six reaction
    are, by definition, unity at zero flow (equilibrium), but are not
    unity for non-zero
    flow $v_p$.
}\label{fig:Hill}
\end{figure}
For the purposes of illustration the thermodynamic constants of the
ten species are $\phi^\std=0$ and $X^\std=1$ and the rate constants of
the six reactions are $\kappa=10$.
This system has a conserved moiety of order 5 involving the the
total amount of the six species ${E}$, ${EM}$, ${LEM}$, ${E^\star}$,
${E^\star M}$ and ${LE^\star M}$ and therefor the method of
\S~\ref{sec:conserved-moieties} is used.
For the purposes of illustration, the total amount is taken as
${E_{tot}=10}$.
Following \citep{GawCra17}, ${M}_o$, ${L}_i$, ${L}_o$ and ${M}_i$ are
chemostats and the amount of ${M}_o$ and ${L}_i$ is taken as unity,
the amount of ${L}_o$ as two and ${M}_i$ is variable.
With these parameters, as derived in the Supplementary Material of
\citep{GawCra17}, the maximum flow is given by
$v_{max}=\frac{100}{21} = 4.76$.

Figure \ref{subfig:Hill_phi} shows flow $v_P$ is plotted against the
variable potential $\phi_{Mi}\si{~mV}$. The maximum flow $v_{max}$ is
marked as a horizontal line.
Figure \ref{subfig:Hill_X_E} shows the amounts of the six species
involved in the conserved moiety; the sum remains 10 but one of the
amounts reaches zero at the maximum flow: this is  why there is a
maximum flow.
Figure \ref{subfig:Hill_rho} shows the values of $\rho_s$
\eqref{eq:MA_sinh} for the six reactions. By definition, $\rho_s=1$ at
zero flow (equilibrium), but is not unity for non-zero flow
$v_p$. Again, one of the $\rho_s$ reaches zero at the maximum flow.


\section{Mitochondrial Electron Transport Chain}
\label{sec:exampl-oxid-phosph}
\begin{figure}[htbp]
  \centering
  \Fig{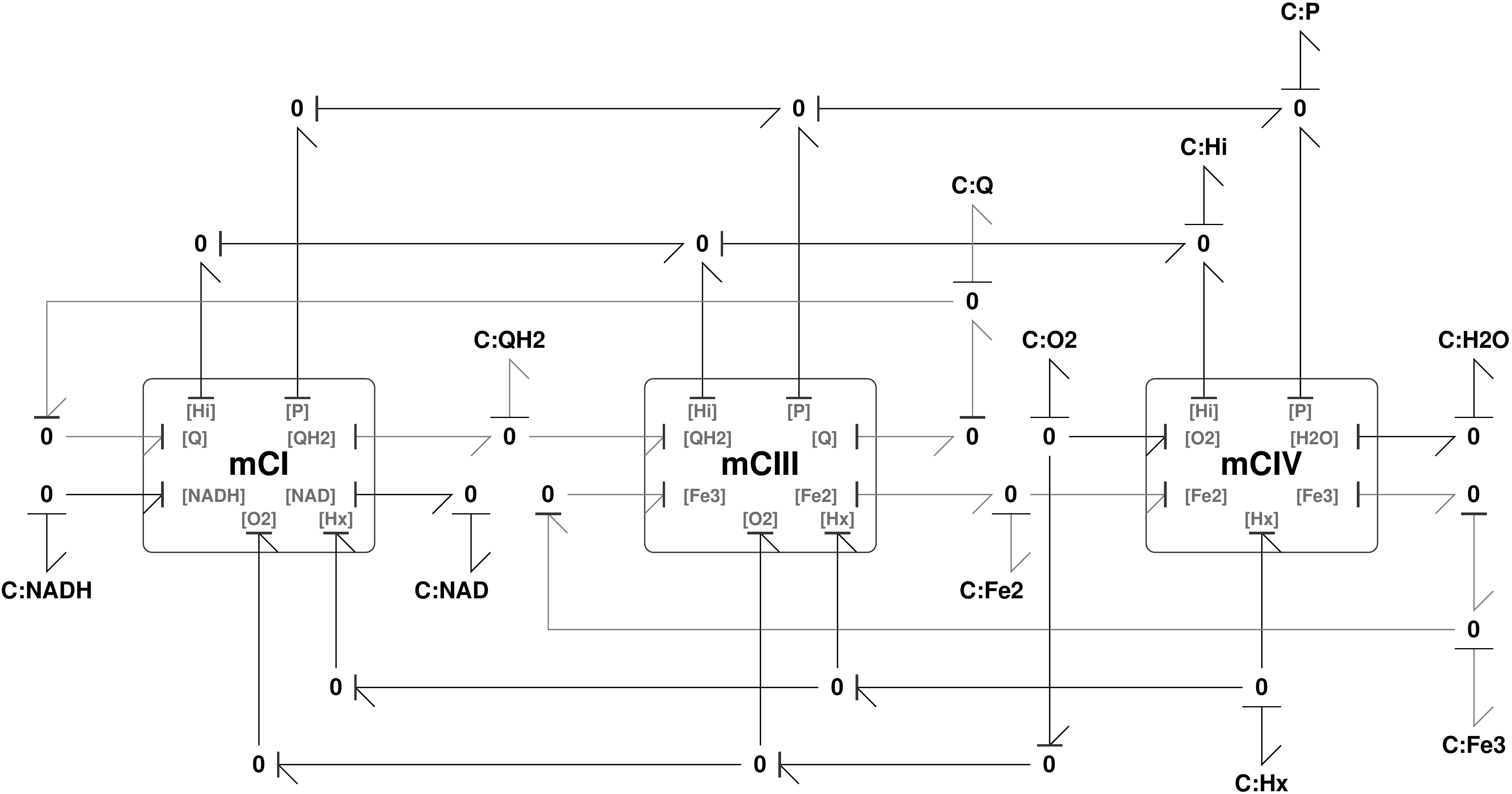}{1}
  \caption{Mitochondrial Electron Transport Chain
    Following \citep{Gaw17a}, the three complexes CI, CIII and CIV
    are represented by the bond graph  modules \textbf{mCI},
    \textbf{mCIII} and \textbf{mCIV} (Figures 5, 6 and 7 of
    \citep{Gaw17a}) interconnected by energy bonds.
}
\label{fig:Ox}
\end{figure}

\begin{figure}[htbp]
  \centering
  \Fig{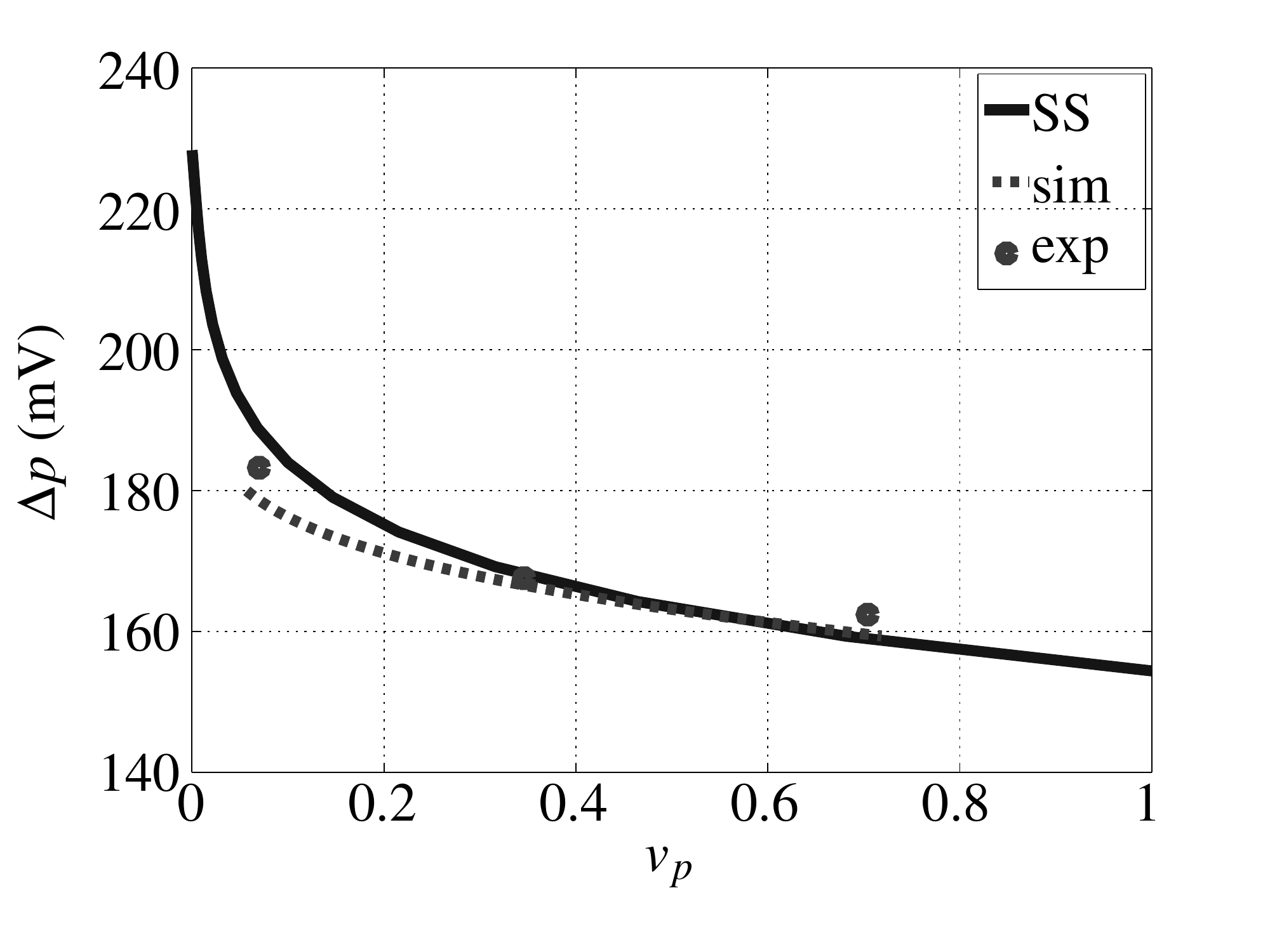}{\FigWidth}
  \caption{Electron Transport Chain: PMF as flow varies.
    The firm line shows the results of steady-state analysis using the
    parameter values discussed in the text. The three points show the
    (normalised) experimetal data of \citet[Figure 2B (low \ch{Pi}
    data)]{BazBeaVin16} and the dashed line the simulation results
    that they present. As discussed in the text, a single parameter was
    adjusted to make the firm line fit the data.  }
  \label{fig:Ox_pmf}
\end{figure}
As discovered by \citet{Mit93}, the key feature of
mitochondria is the \emph{chemiosmotic} energy transduction whereby a
chain of redox reactions pumps protons across the mitochondrial inner
membrane to generate the \emph{proton-motive force} (PMF). This PMF is
then used to power the synthesis of ATP -- the universal fuel of
living systems. The bond graph  model of chemiosmosis of Figure \ref{fig:Ox} is described in detail by
\citet{Gaw17a} and its equilibrium properties analysed.
This paper uses this model to illustrate the steady
state analysis of this biologically significant system and compares
the theoretical results to experimental results of
\citet{BazBeaVin16}.

\citet{BazBeaVin16} compare experimental results with a detailed ode
model of mitochondria which includes not only the Electron Transport
Chain but also the tricarboxylic acid (TCA) cycle and ATP
generation. The model is simulated to steady-state for a variety of
conditions and a number of model outputs are compared with
experimental data. One of these outputs is the membrane potential
$\psi$ which is compared with experimental data in \citep[Figure
2B]{BazBeaVin16} (Two sets of data are given: for low inorganic
phosphate (\ch{Pi}) and high \ch{Pi} -- the low \ch{Pi} data is used
here).  The values of intermembrane and matrix pH were used to compute
the corresponding proton-motive force (PMF).


The bond graph~\citep{Gaw17a} involves 10 chemical species: \ch{NADH},
\ch{NAD}, \ch{O2}, \ch{H2O}, \ch{Hx}, \ch{Hi}, \ch{Q}, \ch{QH2},
\ch{Fe3}, \ch{Fe2},
the transmembrane potential $\psi$ and the electrical charges used in
modelling the redox reactions. \ch{Q} and \ch{QH2} correspond to
oxidised and reduced ubiquinone respectively and \ch{Fe3} and \ch{Fe2}
correspond to oxidised and reduced cytochrome c respectively.
There are 9 reaction components\citep{Gaw17a}: the left \& right half
redox reactions and the proton pump corresponding to each of the three
complexes CI, CIII and CIV.
Following the analysis of \citet{GawCra17}, the positive pathway matrix is:
\begin{equation}
  K_p =
  \begin{pmatrix}
    2&2&4&2&4&4&4&1&4
  \end{pmatrix}^T
\end{equation}


The corresponding reaction rates $\kappa_s$ of Equation
\eqref{eq:MA_sinh_0} are denoted $\kappa_{iL}$, $\kappa_{iR}$ and
$\kappa_{iP}$ coresponding to the left \& right half reactions and the
proton pump of the three complexes \citep{Gaw17a}. For the steady-state
analysis of Figure \ref{fig:Ox_pmf},
$\kappabar_{iL} = \kappa_{iR} = \infty$, $i=1\dots 3$ and
$\kappabar_{2P} = \kappabar_a$,
$\kappabar_{1P} = \kappabar_{3P}  =\infty$ where $\kappabar_a$ is
the sole adjustable parameter.  

As indicated in Figure \ref{fig:Ox_pmf}, when the sole adjustable
parameter $\kappabar_a=0.025$, this simple choice of adjustable
parameters, together with the Faraday-equivalent potentials listed in
\citep[Table 1]{Gaw17a}, gives a steady-state solution consitent with
the experimental data of \citet{BazBeaVin16}.

\section{Conclusion}
\label{sec:conclusion}
The mass-action reaction kinetics equation -- expressed in terms of
forward and reverse reaction affinity -- is rewritten in terms of
reaction affinity and summed reaction affinity. Combining this with
stoichiometric pathway analysis~\citep{GawCra17} leads to an algorithm
for computing steady state potentials corresponding to non-zero
steady-state reaction flows. This is illustrated using introductory
examples, a bond graph model of a biomolecular cycle, and a bond graph
model of the mitochondrial electron transport chain.

Future work will extend the method to include non mass-action kinetics
such as Michaelis-Menten and to handle more than one pathway. It would
also be of interest to examine the relative merits of the $\sinh$ and
$\tanh$ and to analyse and enhance numerical properties of the
approach.

As noted in the Introduction, the approach in this paper is
specifically tailored to the special properties of biomolecular system
bond graphs. It would be interesting to see to what extent this
specific approach could be applied to the general bond graph approach
\citep{Bre84b,BidMarSca07,GonGal11}; similarly it would be interesting
to see how derivative causality appled to biomolecular system bond
graphs \citep{Gaw17} could be combined with the general approach
\citep{Bre84b,BidMarSca07,GonGal11} to deduce structural properties
from the bond graph.

The illustrative example of \S~\ref{sec:exampl-oxid-phosph} just
matches the proton-motive force (PMF) to the
data~\citep{BazBeaVin16}. It would be interesting to use the other
free parameters to match the other types of data analysed by
\citet{BazBeaVin16}.


\section{Acknowledgements}
Peter Gawthrop would like to thank the Melbourne School of Engineering
for its support via a Professorial Fellowship, Edmund Crampin for
help, advice and encouragement and Michael Pan for comments on an
earlier draft.
The constructive input of the reviewers is greatfully acknowledged.


\end{document}